\documentclass[10pt]{article} \oddsidemargin=0in \evensidemargin=0in \textwidth=6.25in
\usepackage{graphicx} 
\begin{document}

\title{Thin films of van der Waals fluid:\\From interface interactions to wetting transitions}

\author{Arik Yochelis \\
Department of Chemical Engineering, \\ Technion -- Israel Institute of Technology, 32000 Haifa, Israel
\and Len M. Pismen \\
Department of Chemical Engineering and \\ Minerva Center for Nonlinear Physics of Complex Systems,
\\ Technion -- Israel Institute of Technology, 32000 Haifa, Israel
}

\maketitle

\begin{abstract}
We present a theoretical study of wetting phenomena and interactions between liquid--vapor
interfaces based on the local density functional theory. The focus is mostly on the impact of
long-range van der Waals interactions both within the fluid and between the fluid and substrate.
For the latter, we consider two models -- hard wall and soft wall approximations -- differing by
the role of steric effects and leading to a qualitatively different character of phase transitions.
We compute numerically the disjoining and conjoining potentials (which are important dynamically
for spreading, spinodal dewetting, and coarsening in thin films, as well as resolution of
interfacial singularities), and loci of intermediate and complete wetting transitions as functions
of the Hamaker constant and temperature. We find that taking into account short-range interactions
is essential for the description of wetting transitions in the soft wall limit. We derive an
analytical form of the disjoining potential and analyze it in the context of the complete,
frustrated and partial wetting.
\end{abstract}

\section{Introduction}

Recent studies of dynamic behavior of thin fluid films and three-phase contact lines showed the crucial
role of precursor layers in contact line dynamics and droplet interaction in the course of spinodal
dewetting and coarsening \cite{pp,pci,pom02}. While the role of precursor layers in wetting and
spreading phenomena had been discussed for a long time \cite{DGen85}, only recently it was realized that
the motion of contact lines is strongly influenced by the precursor layers even when the fluid is not
completely wetting \cite{pp,phf04}.

\textit{Intermediate} wetting transitions leading to the formation of mesoscopic fluid layers, as
distinct from \textit{complete} wetting when a macroscopic layer is formed, were discovered long
time ago~\cite{ebner,NaFi:82,LiKr:84,DiSc:85}, and observed recently experimentally on
solid~\cite{RuTa:92} and liquid~\cite{BKW:92,KBM:93,RMBIB:96,SBRBM:98,RBM:99} substrates. Both
complete and intermediate wetting transitions are assumed to incorporate both short and long-range
interactions. However, a well formulated theoretical study which justifies the individual impact of
such interactions is still missing, though it is recognized as a prime theoretical subject of
interest in view of recent experimental observations and their partial agreement with available
phenomenological models~\cite{BBSRDPBMI:01}.

The mesoscopic dynamic diffuse interface theory~\cite{pp,pci} uses the wide separation between the
width of mesoscopic layers and the scale of hydrodynamic motion to reduce the hydrodynamic
equations to a generalized Cahn--Hilliard equation where the disjoining potential of the fluid
layer~\cite{der} serves as the driving force of the fluid motion along the film. A similar approach
applied to a mesoscopic layer separating two identical bulk fluid phases can be used for the
description of coalescence dynamics, and has been recently discussed in connection to resolving an
interfacial singularity in viscous flow~\cite{cusp}.

The cited works used model expressions for the disjoining potential obtained in the sharp interface
approximation, while emphasizing the essential role of the diffuse character of the vapor-liquid
interface (known already to van der Waals \cite{vdw}) for resolving the interfacial singularities.
The dynamic importance of the disjoining potential in mesoscopic layers necessitates, however,
precise computation linking it to molecular interactions in the fluid and between the fluid and
substrate. The adequate tool for these computations is density functional theory (DFT)
\cite{Eva:79}; further reduction to a local (van der Waals--Landau--Cahn) theory through small
gradient expansion is invalid when the interaction is long-range (i.e. has a power decay
\cite{ditr}). Numerical DFT computations were carried out in the context of phenomenological phase
transition studies \cite{ebner,dagama} without emphasis on computation of disjoining potential.

In this communication, we explore this problem anew using the simplest long-range interaction
potential, which allows for a direct link between the wetting properties and thermodynamics of a
van der Waals fluid. We pay special attention to the character of interactions in the vicinity of
the substrate, distinguishing between ``hard wall'' and ``soft wall'' limits. The first version
used in earlier computations~\cite{dagama} exaggerates steric density depletion near the substrate,
and the opposite limit may be more relevant for rough solid interfaces, as well as for films on
liquid substrates. The character of wetting transitions interactions, as well as properties of the
precursor layer, are very much different in both limits. Realization of the latter reflects the
droplet roll-up behavior under different substrate surfaces; for which the contact angle passes
$\pi$, as was observed in recent experiments~\cite{OSST:96,MSAPN:04}. It is found that the analysis
in the soft wall approximation based solely on van der Waals disagrees with results of recent
experiments on binary fluid systems~\cite{BKW:92,KBM:93,RMBIB:96,SBRBM:98,RBM:99}. We included
therefore a weak dependence on short-range interactions; the expression for the disjoining
potential modified in this way was found to be in a good qualitative agreement with the experiment.

\section{Density functional equations}  \label{S121}

Our starting point is the free energy functional written in the local density functional
approximation~\cite{ll} as
\begin{equation}\label{eq:F}
{\cal F} = \int \rho (\mathbf{r}) f[\rho(\mathbf{r})]\, {\rm d} ^3\mathbf{r}+ \frac{1}{2} \int \rho
(\mathbf{r})\, {\rm d} ^3\mathbf{r} \int_{r'>d} U(r') [\rho (\mathbf{r}+\mathbf{r'})-\rho
(\mathbf{r})] \, {\rm d} ^3\mathbf{r'}, \label{2ff}
\end{equation}
where $f(\rho)$ is free energy per particle of a homogeneous fluid and $U(r')$ is an isotropic pair
interaction kernel with a short-scale cut-off $d$. The functional~(\ref{eq:F}) is written in the
form~\cite{pre01} separating the contribution of density inhomogeneities, expressed by the last
term vanishing in a homogeneous fluid, but is equivalent to an alternative form used in earlier
works~\cite{ditr}.

The chemical potential $\mu =\delta{\cal F}/\delta
\rho$ enters the respective Euler--Lagrange equation obtained by minimizing the grand ensemble
thermodynamic potential $\Phi ={\cal F} - \mu \int \rho \, {\rm d} ^3\mathbf{r}$, which defines the
equilibrium density distribution $\rho (\mathbf{r})$:
\begin{equation}
g(\rho) - \mu + \int_{r'>d} U(r') [\rho
(\mathbf{r}+\mathbf{r'})-\rho (\mathbf{r})] \, {\rm d}
^3\mathbf{r'}=0, \label{2ffg}
\end{equation}
where $g(\rho) ={\rm d} [\rho f(\rho)]/ {\rm d}\rho$. The function $F(\rho)=\rho[ f(\rho)-\mu]$
should have two minima $\rho^\pm$ corresponding to two stable uniform equilibrium states of higher
and lower density (liquid and vapor).

A simple example of long-range potential is the modified Lennard--Jones potential with hard-core
repulsion:
\begin{equation}
U = \left\{ \begin{array}{ccc}
-C_W r^{-6} & {\rm at} & r>d \\
\infty & {\rm at} & r<d
\end{array} \right. \, ,
\label{ljhc}
\end{equation}
where $d$ is the nominal hard-core molecular diameter. The interaction kernel $U(r)$ gives the free
energy density of a homogeneous van der Waals fluid~\cite{pre01}
\begin{equation}
 f(\rho,T) = T \ln \frac{\rho }{1-b\rho} -a \rho,
\label{1ffab}  \end{equation}
where $T$ is temperature, $b=\frac{2}{3}\pi d^3$ is the excluded volume
and
\begin{equation}
a = -2 \pi \int_{d}^\infty U(r)r^2 \, {\rm d} r = \frac{2\pi C_W}{3d^3}. \label{1ff0}
\end{equation}
Equilibrium between the two homogeneous states, $\rho=\rho_0^\pm$ is fixed by the Maxwell condition
\begin{equation}
\mu_0 = \frac{ \rho^+_0  f(\rho^+_0) - \rho^-_0 f(\rho^-_0)} { \rho^+_0  - \rho^-_0}, \label{ff1m}
\end{equation}
which defines, together with $\mu_0 = g(\rho_0^\pm)$, the equilibrium chemical potential
$\mu=\mu_0$ and both equilibrium densities.

The equation for the density distribution near a flat boundary normal to the $z$ axis is obtained
by assuming $\rho$ to be constant in each lateral plane and integrating Eq.~(\ref{2ff}) in the
lateral directions. This yields the free energy per unit area, or surface tension
\begin{equation}
\gamma = \int_{-\infty}^\infty \rho(z) [f(\rho) - \mu]  {\rm d} z +
\frac{1}{2}\int\limits_{-\infty}^\infty \rho(z) \,{\rm d} z \int\limits_{-\infty}^\infty Q(\zeta)
[\rho (z+\zeta)-\rho(z)] \, {\rm d} \zeta. \label{2ff1}
\end{equation}
The interfacial energy is contributed both by deviations from the equilibrium density levels in the
transitional region and by the distortion energy localized there. The 1D interaction kernel $Q(z)$
lumps intermolecular interaction between the layers $z=$ const. It is computed by lateral
integration using as an integration variable the squared distance $q=r^2=\xi^2+z^2$, where $\xi$ is
radial distance in the lateral plane. Taking note that the lower integration limit for $q$ is
$q_0=z^2$ at $|z|>d$, $q_0=d^2$ at $|z| \leq d$, we compute
\begin{equation}
Q(z) = -\pi C_W \int_{q_0}^\infty q^{-3} \, {\rm d} q =
 \left\{
 \begin{array}{ccc} -\frac{1}{2}\pi C_W z ^{-4} & {\rm at} & |z|>d \\
 \\
    - \frac{1}{2}\pi C_W d^{-4} & {\rm at}  & |z|\leq d.
    \end{array} \right.
\label{2ff1a}
\end{equation}

The respective 1D Euler--Lagrange equation, replacing Eq.~(\ref{2ffg}), is
\begin{equation}
g\left [ \rho(z) \right ] - \mu + \int_{-\infty}^\infty  Q(\zeta) [\rho (z+\zeta)-\rho (z)] \, {\rm d}
\zeta =0. \label{2ffg1}
\end{equation}
This equation can be rewritten in a dimensionless form
\begin{equation}
g(\rho) - \mu + \frac{3}{4} \beta \int_{-\infty}^\infty Q(\zeta) [\rho (z+\zeta)-\rho (z)] \, {\rm d} \zeta=0, \label{2ffgwq}\\
\end{equation}
where
\begin{equation}
g(\rho)= \frac{1}{1-\rho} - \ln\left(\frac{1}{\rho}- 1 \right) -   2 \beta\rho . \label{2ffgw}
\end{equation}
Here the length is scaled by the nominal molecular diameter $d$, the density by $b^{-1}$, and the
chemical potential by $T$; the interaction kernel is $Q(z)=-z^{-4}$ at $\vert z \vert>1$, $Q(z)=-1$
at $\vert z \vert \leq 1$, and the only remaining dimensionless parameter is the rescaled inverse
temperature $\beta=a/(bT)$.

\begin{figure}
\begin{center}
\includegraphics[angle=270,width=3in]{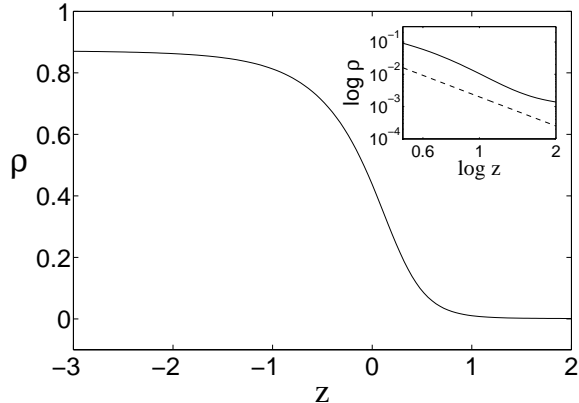}
\end{center}
\caption{The density profile of the liquid-vapor interface at $\beta=9$ obtained by numerical
solution of~(\ref{2ffgwq}). The inset shows the vapor-phase tail of the numerical solution (solid
line) compared to the asymptotic form~(\ref{ffgw1}) depicted by the dashed line.} \label{fig:fdens}
\end{figure}

An example of a density profile obtained by solving numerically Eq.~(\ref{2ffgwq}) is shown in
Fig.~\ref{fig:fdens}. The density tail asymptotics can be estimated by considering a location far
removed from the interface placed at the origin [$|z| \gg 1$ in the dimensionless units of
Eq.~(\ref{2ffgwq})] where a sharp interface limit can be implemented. The density is presented as $
\rho=\rho^\pm_0+ \widetilde \rho$, where $\widetilde \rho/\rho \sim 1/\vert z \vert^3 \ll 1$ .
Inserting this in~(\ref{2ffgw}) and linearizing around $\rho=\rho^\pm_0$, we see that the densities
inside the integral are well approximated in the leading order by the two limiting constants, which
is equivalent to the sharp interface limit. For example, for the vapor tail at $z>0$, $|z| \gg1$ we
have $\rho(z)=\rho^-_0$ and $\rho(z+\zeta)=\rho^+_0$  for $\zeta > |z| $, $\rho(z+\zeta)=\rho^-_0$
for $\zeta < |z| $. Thus, we obtain
\begin{equation} \label{ffgw1}
\rho =\rho_0^\pm + \frac{\beta(\rho^+_0 - \rho^-_0) }{4 g\,'(\rho_0^\pm)}\frac{1}{z^3} \,.
\end{equation}
This is in good agreement to the numerical solution, as seen in the inset of Fig.~\ref{fig:fdens}.
One can check \emph{a posteriori} using this expression that the contribution to the integral of
neighboring locations with $|\zeta| =O( 1)$ is of a higher order $\propto| z|^{-5}$ and therefore can be neglected.

\section{Interacting fluid-vapor interfaces} \label{S14}

If two flat fluid-vapor interfaces are in close proximity, the equilibrium chemical potential and
surface tension become dependent on their mutual separation $h$. This phenomenon is relevant for
processes of topology change, e.g.\ droplet coalescence. The corrections due to proximity of
interfaces in the case when a gas layer separates two identical semi-infinite bulk liquids can be
obtained by finding a homoclinic solution of Eq.~(\ref{2ffgwq}) satisfying the asymptotic
conditions $\rho(\pm \infty)=\rho^+$. A stationary solution of such kind exists at certain values
of $\mu$ shifted downwards from the Maxwell construction level $\mu_0$. The shift $\Delta
\mu_c=\mu_0-\mu$ corresponds to the \emph{conjoining potential} expressing the interaction of two
identical flat interfaces.

A rough but qualitatively correct approximation can be obtained by computing molecular interactions
between two sharp interfaces~\cite{cusp}. The shift of chemical potential necessary to keep two
interfaces separated by a distance $h$ in the state of equilibrium is determined in this
approximation by the decrement of the integral in Eq.~(\ref{2ffgwq}) due to replacing gas by liquid
at $z>h$. For the purpose of a qualitative estimate, the sharp-interface computation valid at $h
\gg 1$ can be extended also to small separations, ($h \le 1)$ to find~\cite{cusp}
\begin{equation}
\Delta \mu_c
 = \left\{ \begin{array}{lll}
 \frac{4}{3}-h
& \mathrm{at} &h \leq 1 ,  \\
& \\
    \frac{1}{3} h^{-3}
& \mathrm{at} & h > 1 , \end{array} \right. \label{gamma}
\end{equation}
Equilibrium of a layer between two interfaces is unstable; the instability is, however, very weak
when separation of interfaces is large compared to the molecular scale. Localized small
perturbations decay under these conditions due to surface tension, and a large disturbance is
needed to create a critical nucleus initiating the topology change.

\begin{figure}
\begin{center}
\includegraphics[angle=270,width=3in]{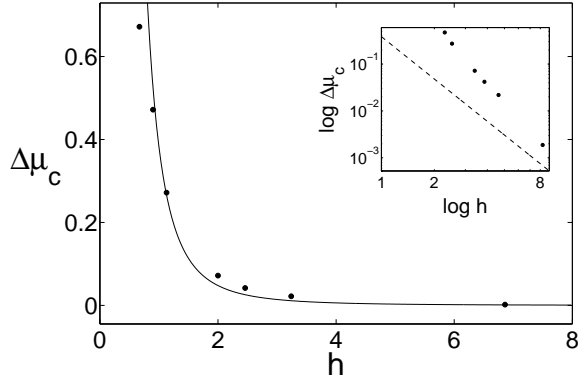}
\end{center}
\caption{{The dependence of the dimensionless conjoining potential $\Delta \mu_c$ on separation
$h$. The dots denote the results of 1D density functional computation with a shift of
$h_\star=1.39$ adjusted to fit Eq.~(\ref{gamma}), as shown by the solid line. Inset: the tail of
the numerical dependence $\Delta \mu_c(h)$ compared to the $h^{-3}$ decay (dashed line).}}
\label{fmuh}
\end{figure}

A precise dependence is obtained by solving numerically Eq.~(\ref{2ffgwq}). The solution is found by
fixing some trial value of $\mu$ and solving Eq.~(\ref{2ffgwq}) iteratively to find a stationary profile
$\rho(z)$ at this value. The nominal gap width is defined as
\begin{equation}
h =  \frac{1}{\rho^+-\rho^-} \int_{-\infty}^\infty
  \left(\rho^+-\rho\right) {\rm d} z .
\label{2inth}  \end{equation}
The computation results for $\beta=9$ are shown by dots in
Fig.~\ref{fmuh}. The curve $\Delta \mu_c(h)$ expressing this dependence well fits the computational
results shifted by a certain value $h_\star$, equal to $\approx 1.39$ in this particular
computation. A shift is necessary because, while the separation in Eq.~(\ref{gamma}) can be formally
arbitrarily small, no stationary solution of Eq.~(\ref{2ffgwq}) can exist below a certain value of
$h$ which corresponds to a critical size required for nucleation of a critical 1D ``bubble''. The applied shift
equals to the width of this ``bubble'' computed according to Eq.~(\ref{2inth}).

\section{Fluid-substrate interactions}

\subsection{Disjoining potential} \label{S15}

In the proximity of a substrate surface, the additional term in the free energy integral
(\ref{2ff}) is
\begin{equation}
{\mathcal F}_s = \int \rho(\mathbf{r}) \,{\rm d} ^3\mathbf{r} \int_s U_s(\vert r-r' \vert) \rho_s
(\mathbf{r'}) \, {\rm d} ^3\mathbf{r'}\, , \label{ffs}
\end{equation}
where $ U_s$ is the attractive part of the fluid-substrate interaction potential, $\rho_s$ is the
substrate density, and $\int_s$ means that the integration is carried over the volume occupied by
the substrate; all other integrals here and in (\ref{2ff}) are now restricted to the volume
occupied by the fluid.

In the following, we shall consider a flat interface parallel to the substrate surface $z=0$, and
suppose that liquid-substrate interactions are also of the van der Waals type with a modified
constant $C_{S}=\alpha_s C_W$. Then the free energy per unit area is expressed, after some
rearrangements, as
\begin{equation}
\gamma_s = \int_0^\infty \rho(z) \left\{f(\rho) + \psi_l(z) \left[\alpha_s \rho_s -\frac{1}{2}
\rho(z) \right] \right\} {\rm d} z + \frac{1}{2}\int_0^\infty \rho(z)\, {\rm d} z\int_{0}^\infty
Q(z-\zeta) [\rho (\zeta)-\rho (z)] \, {\rm d} \zeta.
 \label{ffsv}
\end{equation}
The first term contains the same local part as in Eq.~(\ref{2ff1}) complemented by the
liquid-substrate interaction energy. The latter is computed by integrating the attracting part of
the fluid-fluid and fluid-substrate interaction energy laterally as in Eq.~(\ref{2ff1a}) and
represents the shift of energy compared to the unbounded fluid. The term $\rho(z)/2$ compensates
lost fluid-fluid interactions in the substrate domain which are included in the homogeneous part
$f(\rho)$.

\begin{figure}
\begin{center}
\includegraphics[width=2.15in]{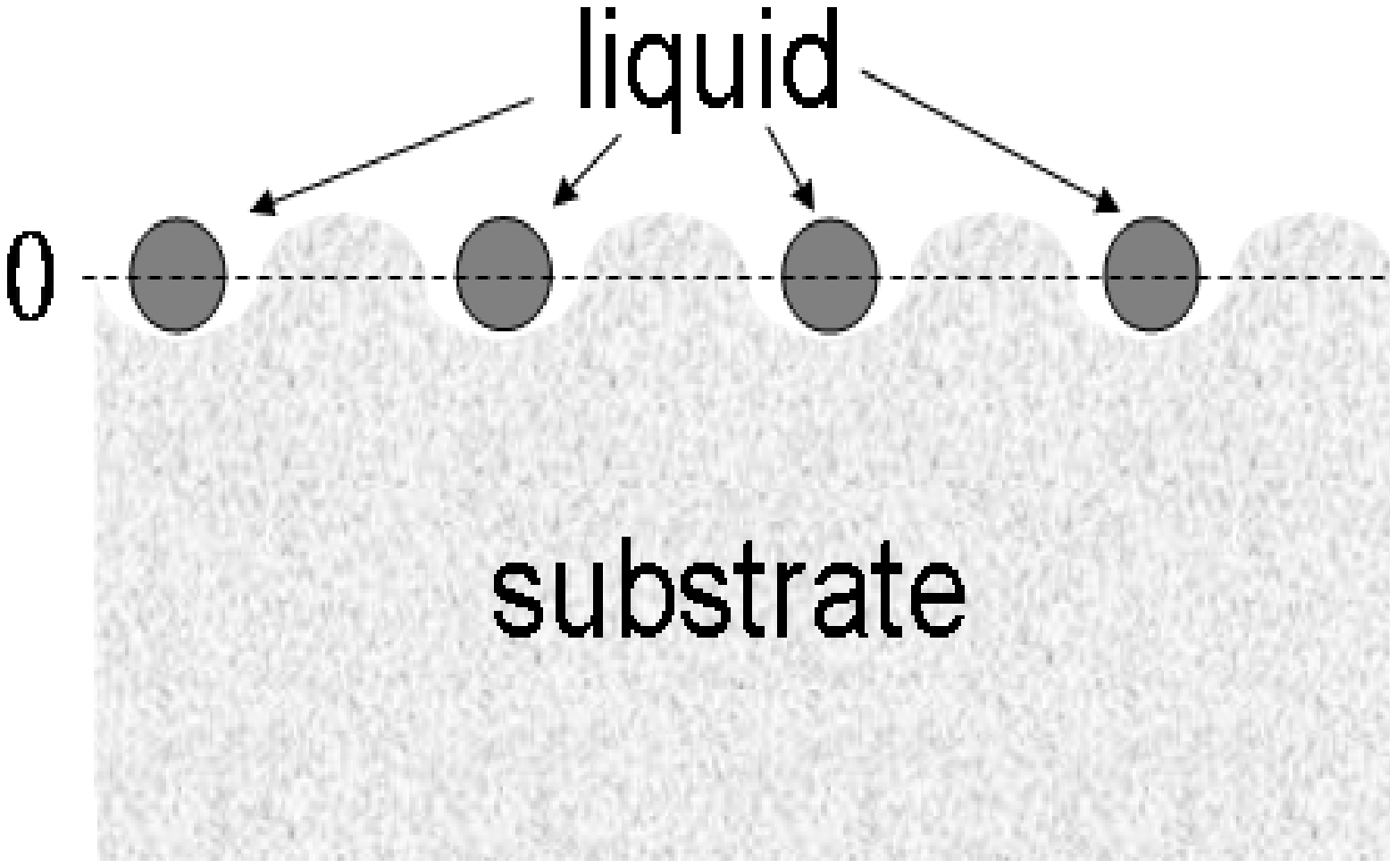}~~~~~~~~~~~
\includegraphics[width=2.15in]{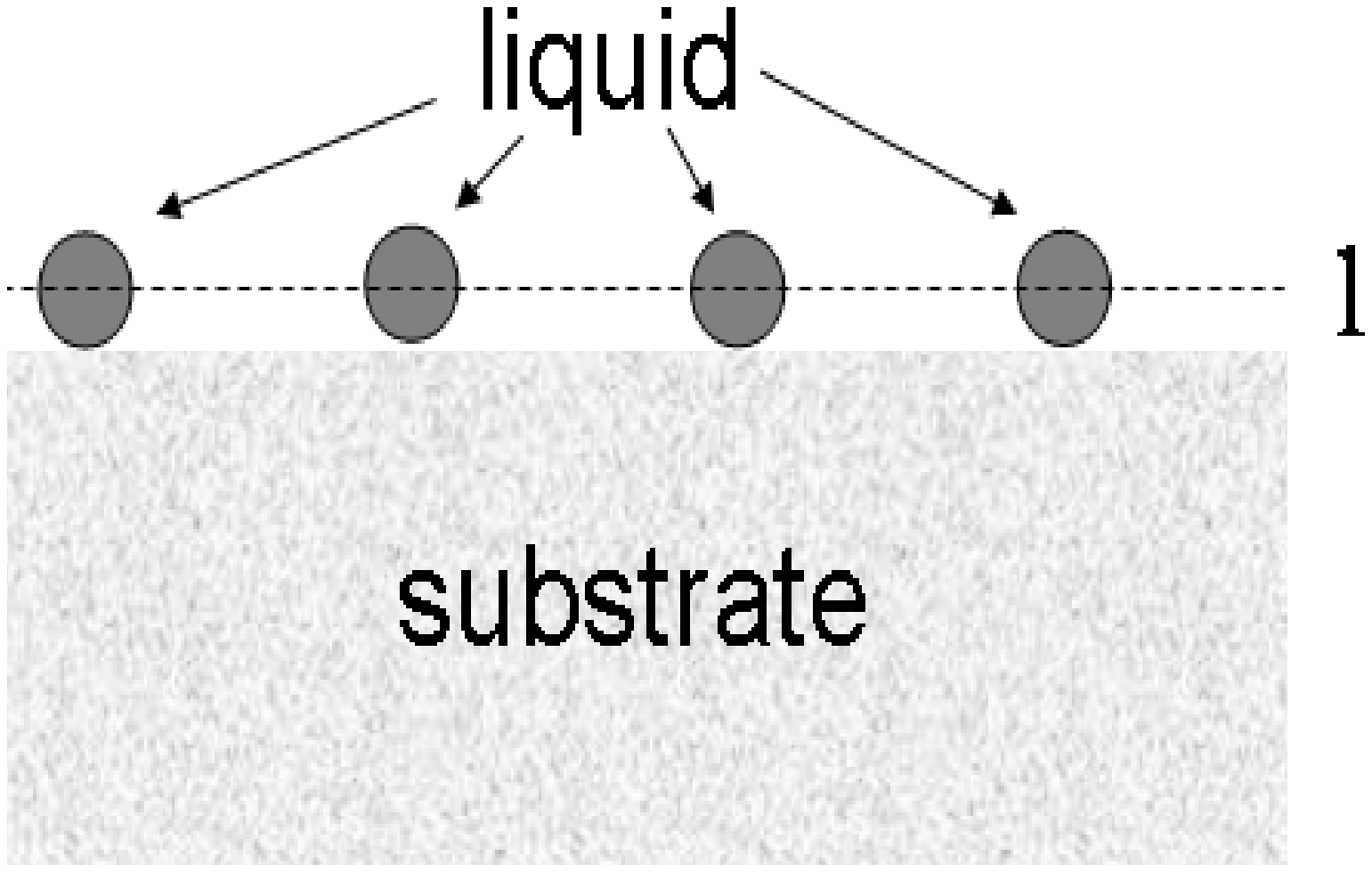}
\end{center}
  \caption{Schematic cartoon representation of soft (left) and hard (right) wall approximations.
  Solid circles represent hard sphere fluid molecules and the horizontal coordinate, in this setup is $z$.}
\label{fig:S-HWA}
\end{figure}

Computation of the function $\psi_l(z)$ depends on steric constraints imposed upon fluid molecules
in the vicinity of the substrate. We consider two limiting cases, as demonstrated in
Fig.~\ref{fig:S-HWA}:
\begin{description}
    \item[(i)] \textit{Soft wall approximation} (SWA) which allows for fluid molecules to
    penetrate up to the surface of the substrate.
    \item[(ii)] \textit{Hard wall approximation} (HWA) which imposes steric constraints
    preventing the centers of fluid molecules from approaching the substrate at distances shorter
    than the hard-core cutoff $d$.
\end{description}
To be definite, we place the origin of the coordinate system on the centerline of the first row of
the substrate atoms. In the soft wall approximation, the computation yields
\begin{equation}
\psi_l(z) = -\pi C_W \int_{-\infty}^0 {\rm d} \zeta
   \int_{q_0}^\infty q^{-3}\,{\rm d} q =  \int_{-\infty}^0 Q(z-\zeta) {\rm d} \zeta,
\label{ff2s}
\end{equation}
where the integration limit is $q_0=(z-\zeta)^2$ at $|z-\zeta|>d$,
$q_0=d^2$ at $|z-\zeta| \leq d$. The result is
\begin{equation}
\psi_l (z) = \left\{ \begin{array}{ccc} -\frac{1}{6}\pi C_W z^{-3} & {\rm at} & |z|>d \\
\\
-\pi C_W d^{-3} \left( \frac{2}{3} - \frac{z}{2d} \right) & {\rm at}  & |z| \leq d .
\end{array}\right.
\label{ff1s}
\end{equation}
The dimensionless Euler--Lagrange equation derived from
Eq.~(\ref{ffsv}) reads
\begin{equation}\label{ffgw_s}
 g(\rho) - \mu + \frac{3}{4} \beta {\psi}_l(z) \left [ \rho^+(\chi+1)
-\rho(z)\right ] + \frac{3}{4} \beta \int\limits_{0}^\infty  Q(\zeta-z) [\rho (\zeta)-\rho (z)] \,
{\rm d} \zeta =0
\end{equation}
where $\chi=\alpha_s \rho_s/\rho^+-1$ is the dimensionless Hamaker constant, and ${\psi}_l(z) =
-z^{-3}/3$ at $z>1$, ${\psi}_l(z) = z-4/3$ at $z\leq 1$.

\begin{figure}
\begin{center}
\includegraphics[angle=270,width=3in]{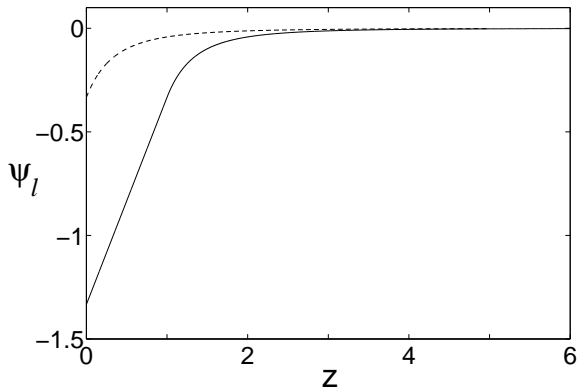}
\end{center}
  \caption{The function ${\psi}_l (z)$ for soft (solid line) and hard (dashed line) wall approximations.}
\label{fig:psi_z}
\end{figure}

The hard wall approximation introduces, in effect, a void layer with the thickness equal to the
hard-core cutoff, hence ${\psi}_l(z)=0$ at $0<z\leq 1$ \cite{dagama}. If the closest allowed
position of the centers of fluid molecules, is taken as the origin, then we define a hard wall
fluid-substrate interaction as ${\psi}_l^{HWA}(z)={\psi}_l^{SWA}(z+1)$ as seen in
Fig~\ref{fig:psi_z}. This shift significantly changes the equilibrium solution and the character of
the wetting transition.

The equilibrium chemical potential is shifted from the Maxwell construction, $\mu=\mu_0$ in the
proximity of the substrate surface. The shift $\Delta \mu_d=\mu-\mu_0$, called  \emph{disjoining
potential}~\cite{der}, can be defined as
\begin{equation}
\Delta\mu_d=\frac{1}{\rho_0^+-\rho_0^-} \, \frac{\partial \gamma_s }{\partial h} , \label{mufs}
\end{equation}
where $h$ is the nominal distance between gas-liquid and liquid-substrate interfaces. The
latter is defined, analogous to Eq.~(\ref{2inth}), as
\begin{equation}
h =  \frac{1}{\rho^+-\rho^-} \int_{0}^\infty
  \left(\rho-\rho^-\right) {\rm d} z .
\label{2inths}  \end{equation}

\subsection{Equilibrium solutions}\label{sec:sol}

The sign of $\chi$ distinguishes a perfectly wetting fluid at $\chi>0$ and a ``nonwetting'' fluid
forming a finite contact angle at $\chi<0$. In the latter case, there are two branches of
equilibrium solutions of Eq.~(\ref{ffgw_s}) parametrized by the disjoining potential $\Delta
\mu_d$. The stable branch with small $h$ is characterized by a monotonic density profile and
corresponds to the vapor phase thickening to a relatively dense adsorbate or precursor layer near
the substrate. The unstable branch with larger $h$ is characterized by a non-monotonic density
profile and corresponds to a liquid layer with a slightly depleted density near the substrate.
Instability is characteristic to any layer of a nonwetting fluid, but it is very weak when
separation of the vapor-liquid and liquid-substrate interfaces is large compared to the molecular
scale. The contact angle can be expressed through the disjoining potential for the case $|\chi| \ll
1$ when a small-angle approximation is valid \cite{pre01}:
\begin{equation}\label{eq:cont_angl}
    \theta=\sqrt{\frac{2\rho^+}{\gamma}\int^\infty_{h_0}\Delta \mu_d {\rm d}h}\,,
\end{equation}
where $h=h_0$ is the precursor layer thickness defined by the condition $\mu(h_0)=\mu_0$
or $\Delta \mu_d(h_0)=0$.

\begin{figure}
\begin{center}
  \includegraphics[angle=270,width=3.1in]{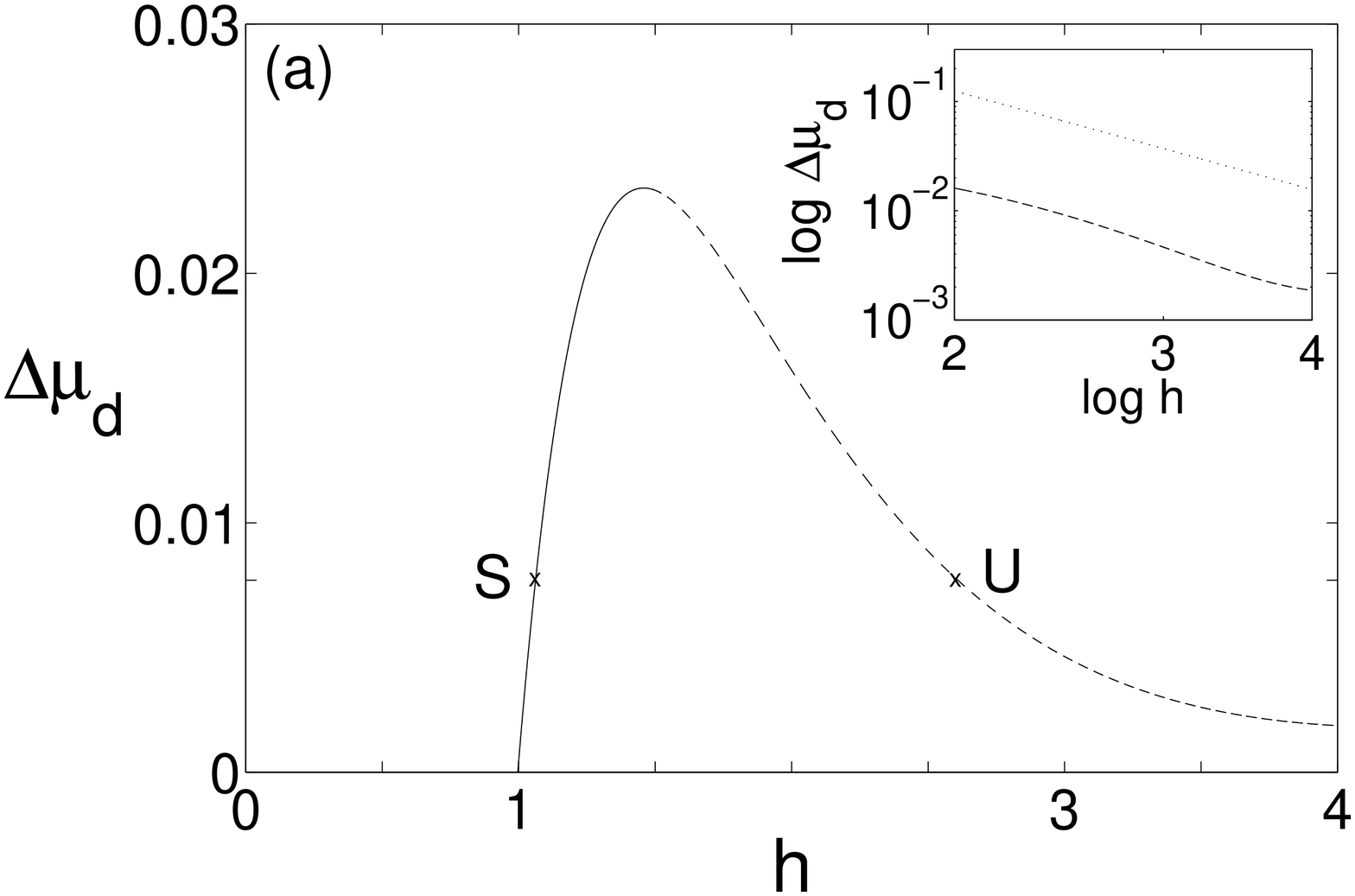}
  \includegraphics[angle=270,width=3in]{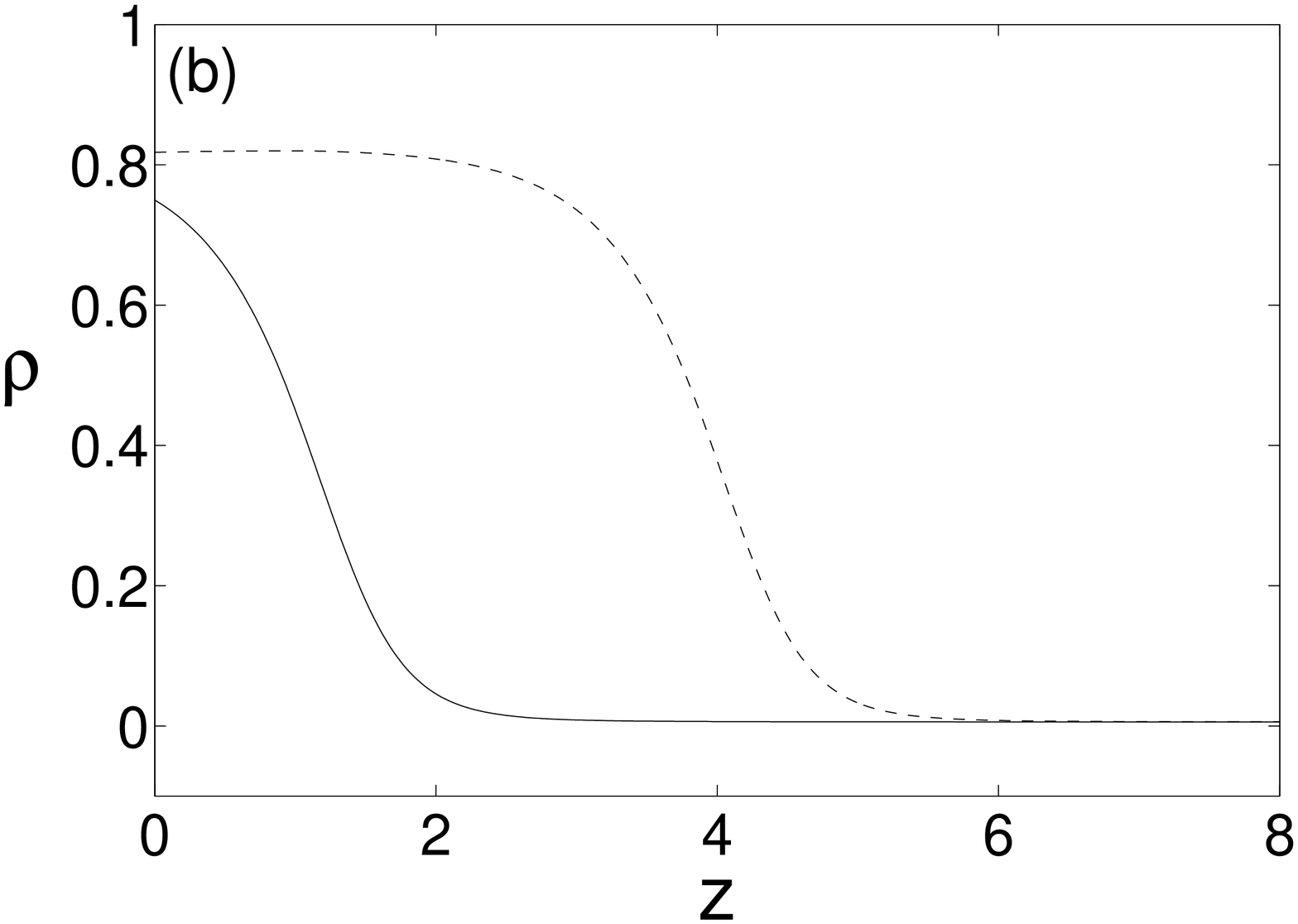}
\end{center}
  \caption{Equilibrium solutions of Eq.~(\ref{ffgw_s}) in the soft wall approximation (SWA).
  (a) The dependence of the dimensionless disjoining potential $\Delta \mu_d$ on separation $h$.
  The solid/dashed lines denote stable/unstable solutions, respectively. The dotted line in the
  inset depicts the $h^{-3}$ decay. (b) Coexisting density profiles at $\Delta \mu_d=2.2\cdot10^{-3}$.
  Stable (solid line) and unstable (dashed line) profiles correspond to $S$ and $U$ in (a) respectively.
  Parameters: $\chi=-0.05$, $\beta=7$.
  }
\label{fig:slg_S}
\end{figure}

\begin{figure}
\begin{center}
  \includegraphics[angle=270,width=3.1in]{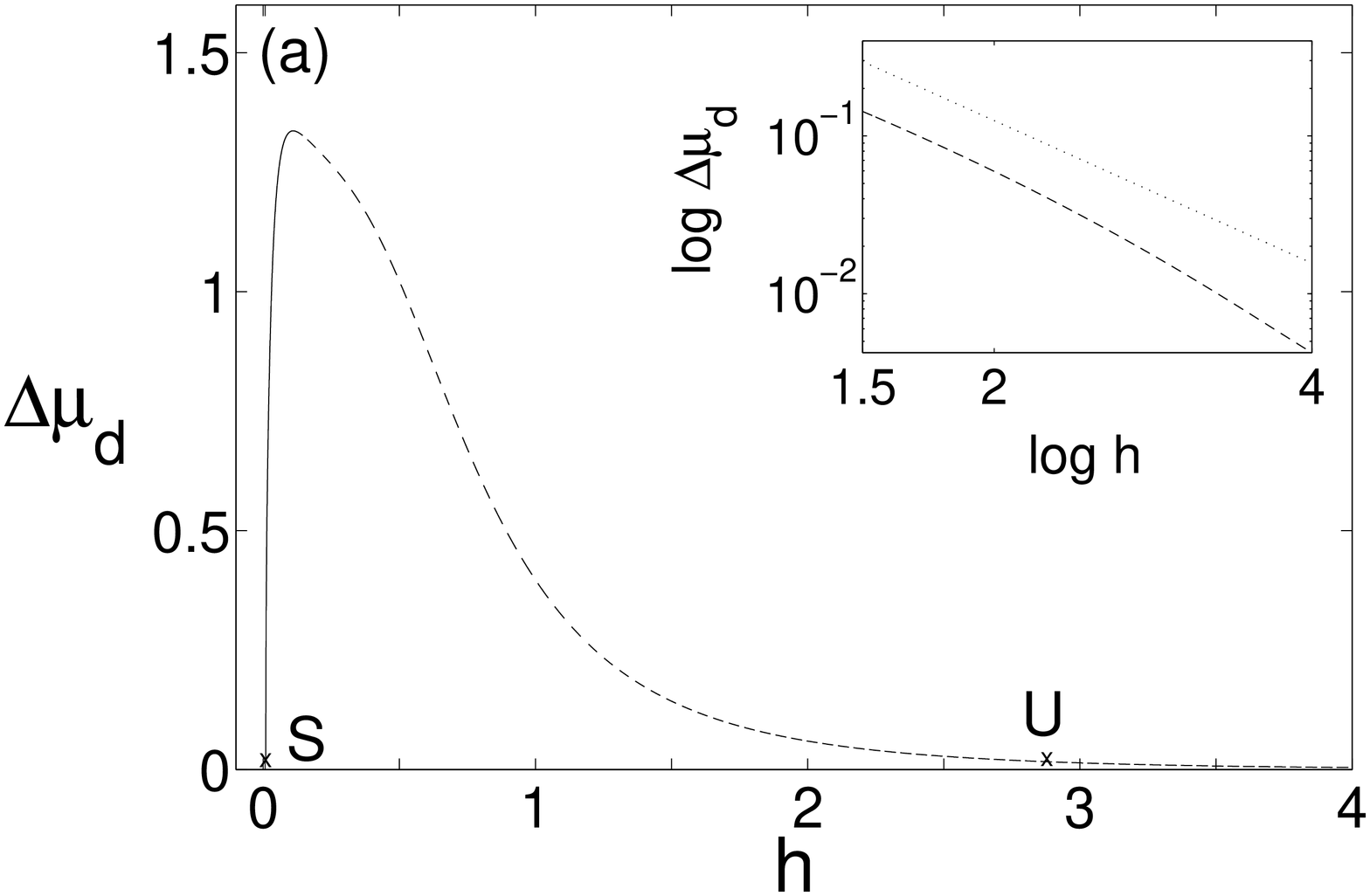}
  \includegraphics[angle=270,width=3in]{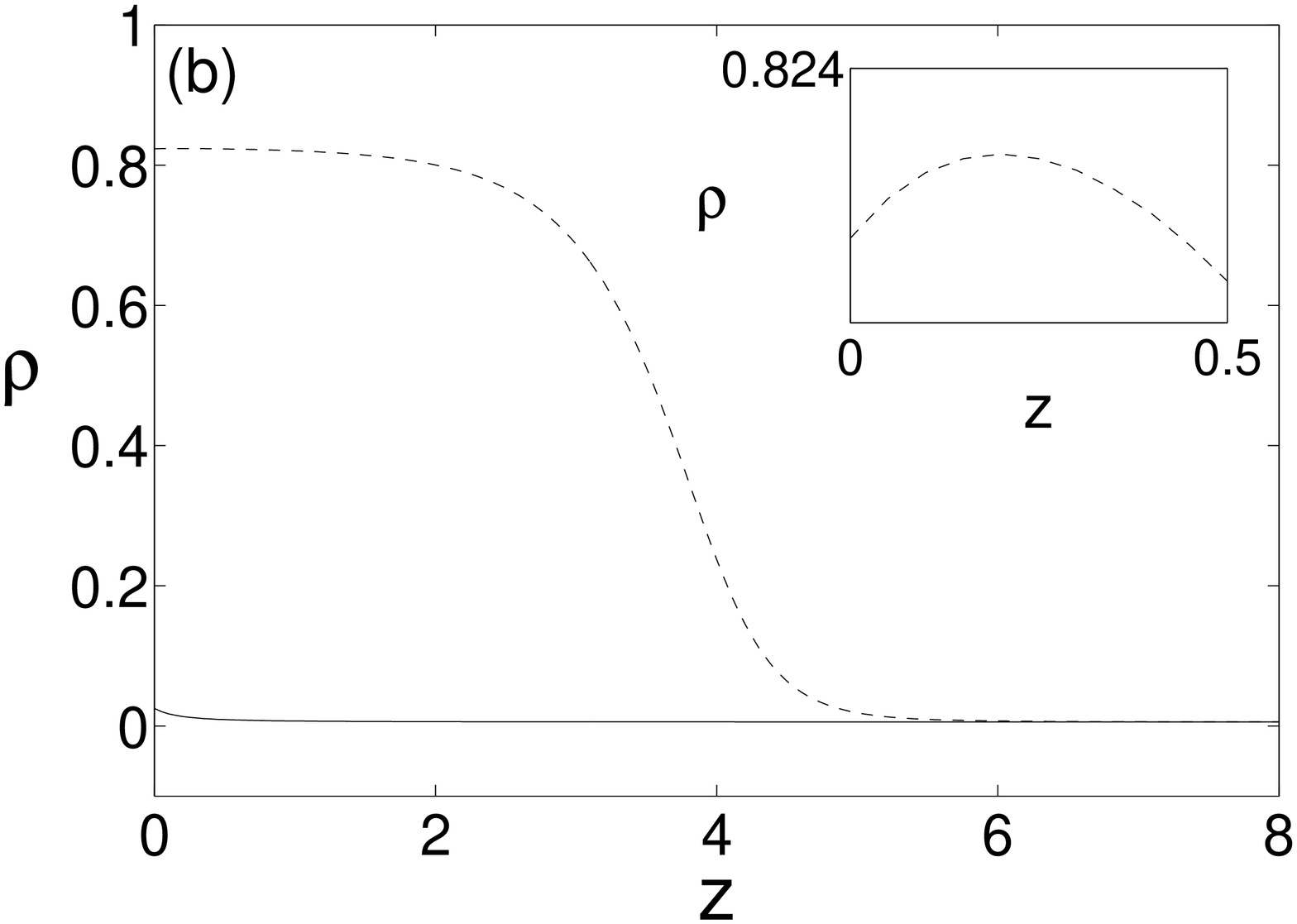}
\end{center}
 \caption{Equilibrium solutions of Eq.~(\ref{ffgw_s}) in the hard wall approximation (HWA).
  (a) The dependence of the dimensionless disjoining potential $\Delta \mu_d$ on separation $h$.
  The solid/dashed lines denote stable/unstable solutions, respectively. The dotted line in the inset
  depicts the $h^{-3}$ decay. (b) Coexisting density profiles at $\Delta \mu_d=7.7\cdot10^{-3}$.
  Stable (solid line) and unstable (dashed line) profiles correspond to $S$ and $U$ in (a) respectively.
  Parameters: $\chi=-0.05$, $\beta=7$.
  }
\label{fig:slg_H}
\end{figure}

Figures~\ref{fig:slg_S}(a) and~\ref{fig:slg_H}(a) present typical equilibrium curves  $\Delta
\mu_d(h)$, respectively, for the SWA and HWA. Examples of the corresponding density profiles are
shown in Fig.~\ref{fig:slg_S}(b) and~\ref{fig:slg_H}(b). One should keep in mind that in the hard
wall approximation, $h$ should be rescaled back so that the density profiles start at $h=1$. Unlike
in the soft wall approximation, in the hard wall case, the void between the substrate and the
liquid density encourages density depletions [see inset of Fig.~\ref{fig:slg_H}(b)]. All solutions
exhibit a $\Delta \mu_d \sim h^{-3}$ tail at large $h$, in agreement to calculations performed in
the sharp interface approximation~\cite{pre01}. Oscillatory density tails cannot appear in our
model, unlike more sophisticated nonlocal DFT computations~\cite{henderson,dagama}. Although the
curves $\Delta \mu_d(h)$ are qualitatively similar to those obtained in the sharp interface
approximation \cite{pre01}, the quantitative distinctions strongly influence the character of the
wetting transition, as will be emphasized in the following.

\subsection{Comparison between HWA and SWA}

Investigating numerically the two above substrate-liquid interaction models we can distinguish
between two main differences in the interaction properties: the emergence of  ``microscopic''
solutions, identified with nanoscale  precursor layers and transition to layers of mesoscopic or
macroscopic thickness. The three classes of films correspond, respectively to $h \sim {\mathcal
O}(1)$, $h \sim {\mathcal O}(10 )$, $h >{\mathcal O}(10^2 )$  (measured on the molecular scale
$d$).

\subsubsection{Precursor layer}

\begin{figure}
\begin{center}
\includegraphics[angle=270,width=2.9in]{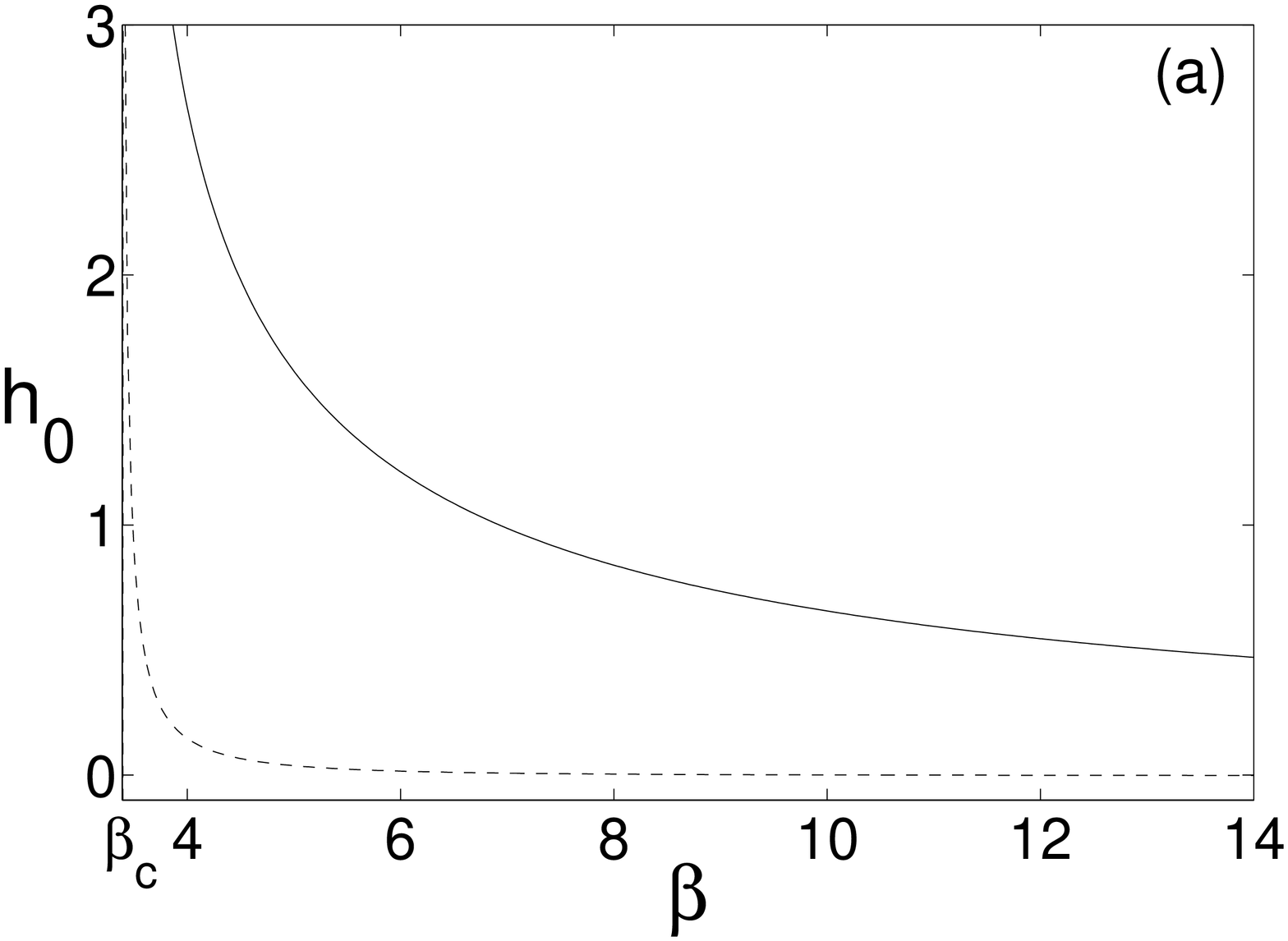}
\includegraphics[angle=270,width=3.08in]{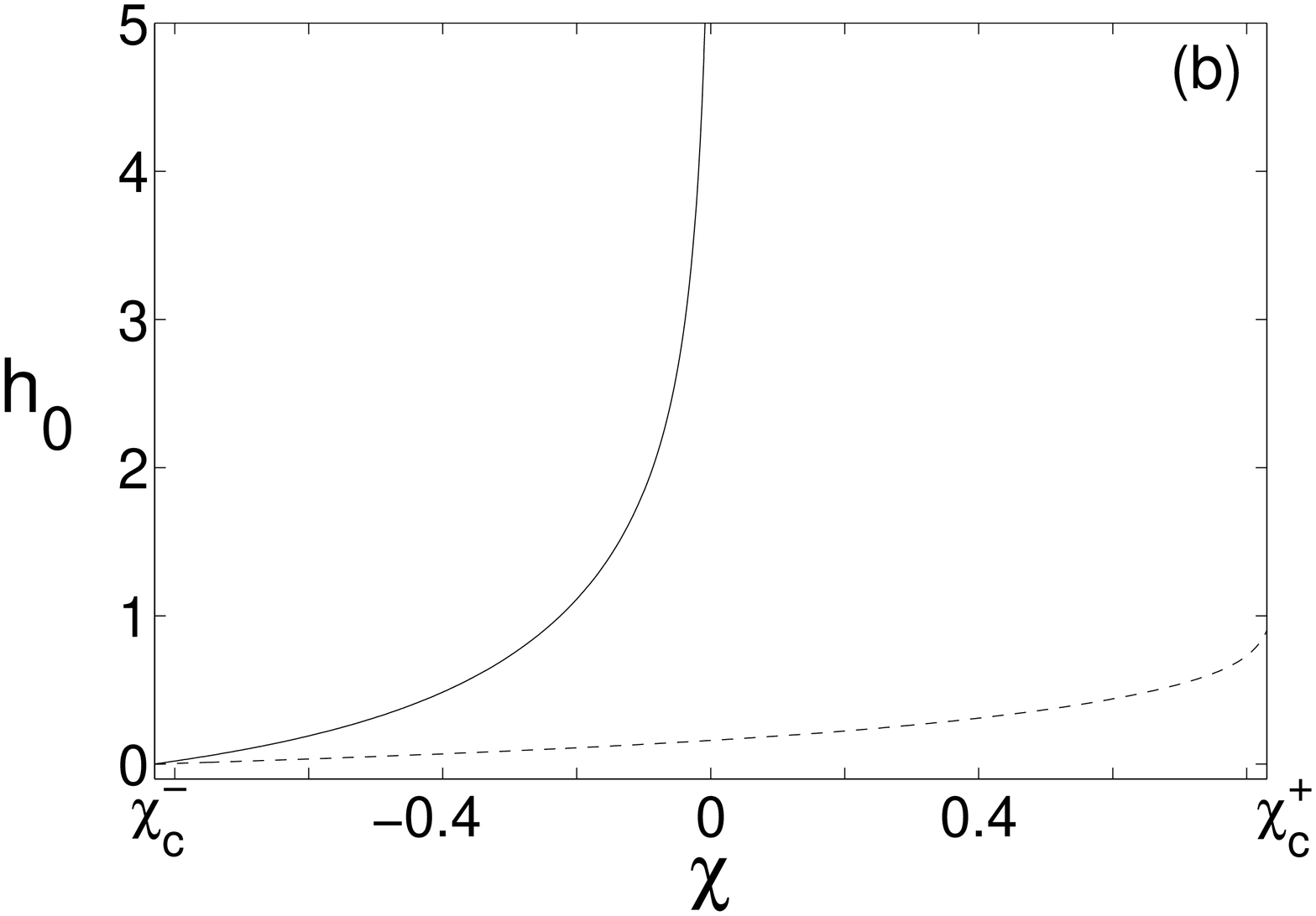}
\end{center}
  \caption{The dependence of the dimensionless precursor layer thickness on (a) the inverse temperature
  $\beta$ for a fixed value of the dimensionless Hamaker constant $\chi=-0.05$; and (b) the Hamaker constant
  $\chi$ at $\beta=4$. The solid/dashed lines represent the soft/hard wall approximations,
  respectively. In (b) $\chi_c^-$ corresponds to the emergence threshold of the precursor layer,
  identical in both SWA and HWA models. The precursor thickness at the wetting transition threshold
  $\chi=\chi_c^+$ is finite in HWA, indicating a first order transition, while in SWA $h_0\rightarrow\infty$ at $\chi \rightarrow \chi^+_c=0$, indicating a  second order transition. } \label{fig:beta_h0}
\end{figure}

Stable equilibrium solutions with a finite thickness $h=h_0$, which correspond to a microscopic
precursor layer, exist at the liquid-vapor equilibrium chemical potential $\mu=\mu_0$. An example
of the dependence of the dimensionless precursor layer thickness on the inverse temperature $\beta$
is shown in Fig.~\ref{fig:beta_h0}(a). One can see a strong difference between SWA and HWA results.
In the HWA computation, the value of $h_0$ at the chosen value of the Hamaker constant is much less
than unity, unless near the critical temperature $\beta=\beta_c \simeq 3.37$, so that one can speak
of a dilute adsorption layer rather than of a proper precursor. This difference stems from an
effective increase of the absolute value of the Hamaker constant due to the presence of a dilute
layer of steric origin present in HWA.

\begin{figure}
\begin{center}
  \includegraphics[angle=270,width=3.15in]{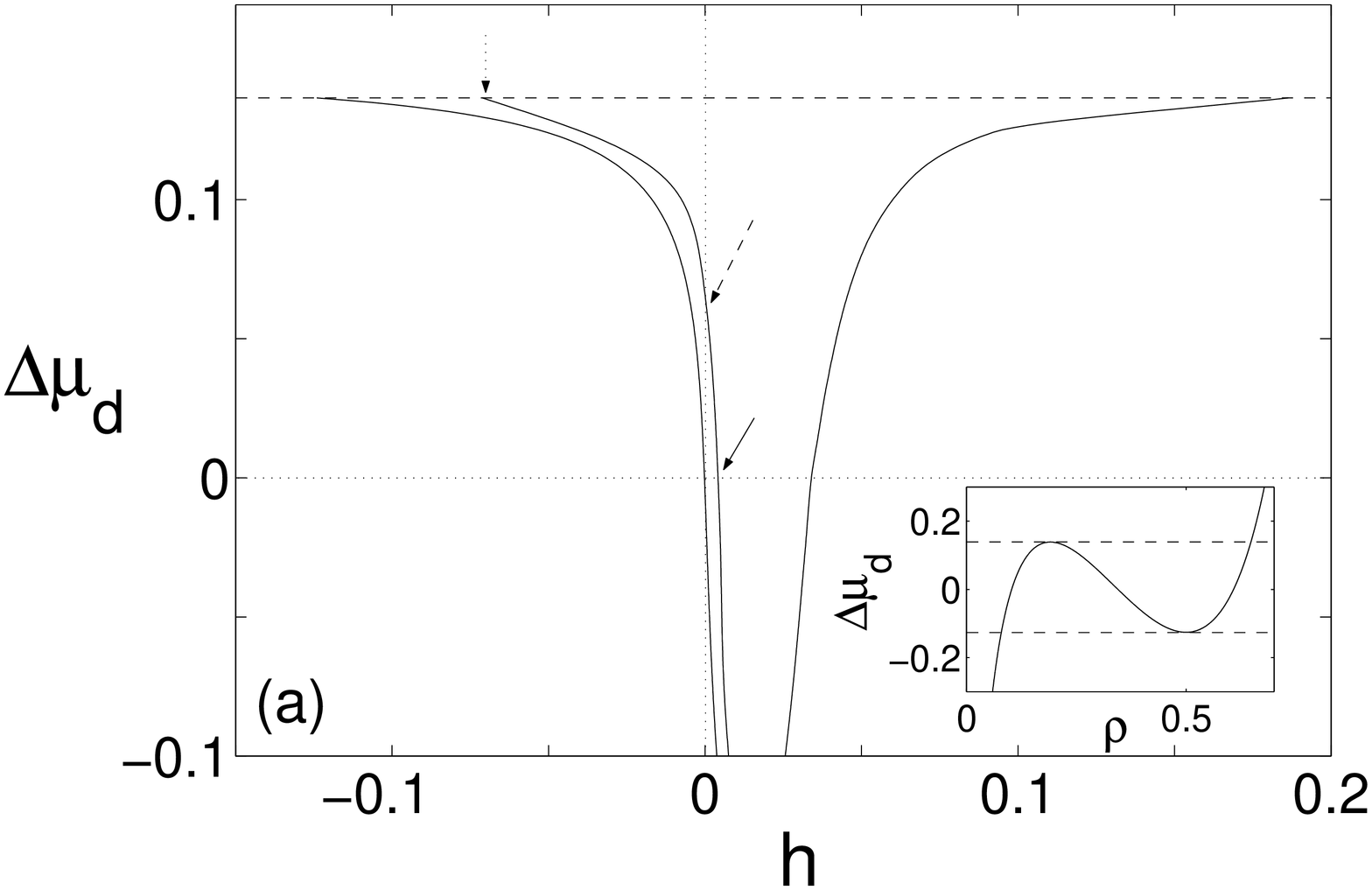}
  \includegraphics[angle=270,width=3in]{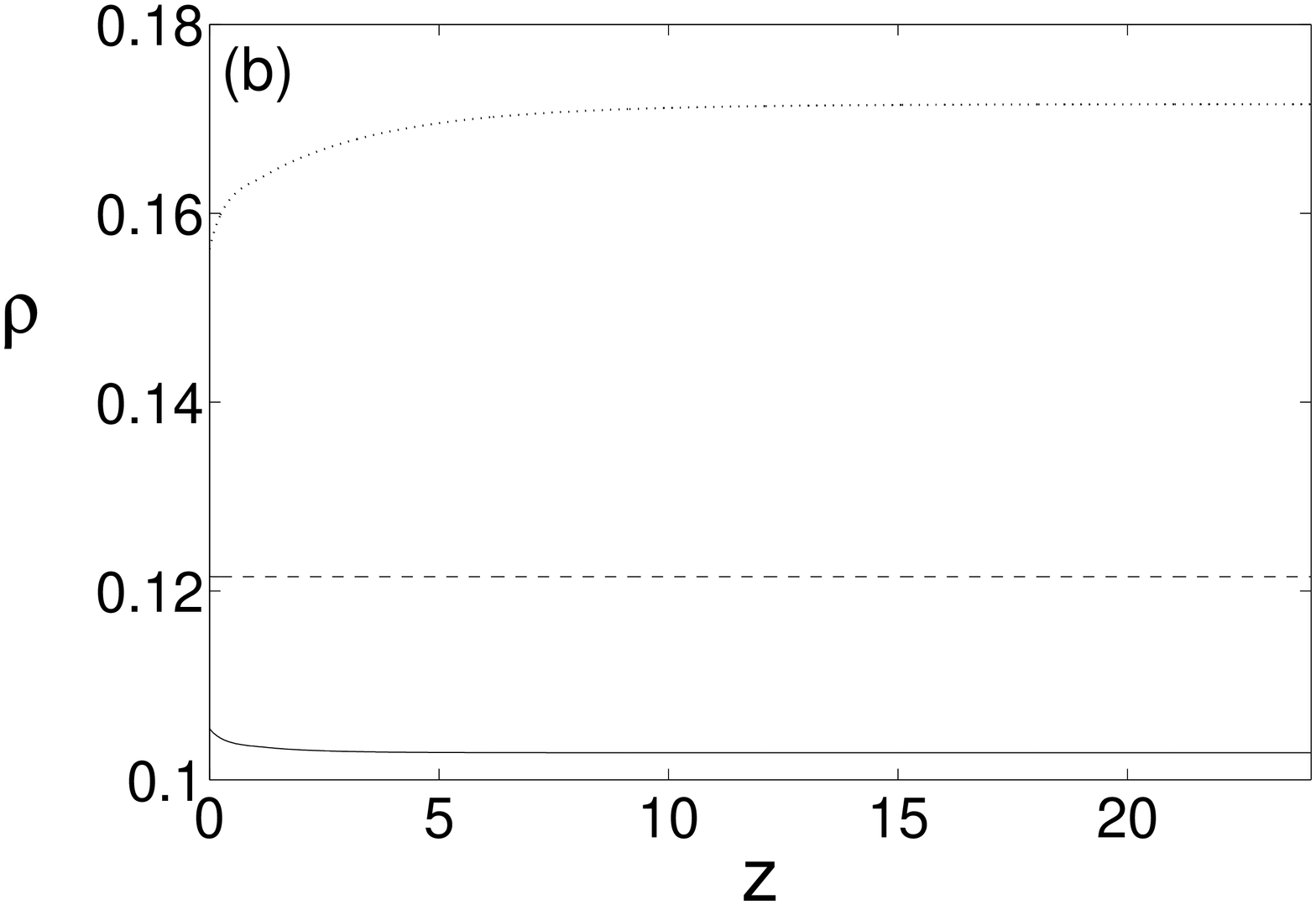}
\end{center}
  \caption{Equilibrium solutions of Eq.~(\ref{ffgw_s}) in the hard wall approximation (HWA) in the
  vicinity $\chi=\chi^-_c=-0.833$.
  (a) The dependence of the dimensionless disjoining potential $\Delta \mu_d$ on separation $h$ at $\beta=4$ and
  different values of the Hamaker constant, from right to left, $\chi=-0.6,\,-0.8,\,\chi_c^-$. The dashed line represents
  the upper limit of the vapor-liquid coexistence. The coexistence range is demonstrated in the inset.
  The dashed line denotes the critical shift of chemical potential at which the curve $\Delta \mu_d(h)$ becomes discontinuous.
  (b) Density profiles corresponding to the arrows in (a).
  }
\label{fig:slg_H_den}
\end{figure}

\begin{figure}
 \begin{center}
 \includegraphics[angle=270,width=3in]{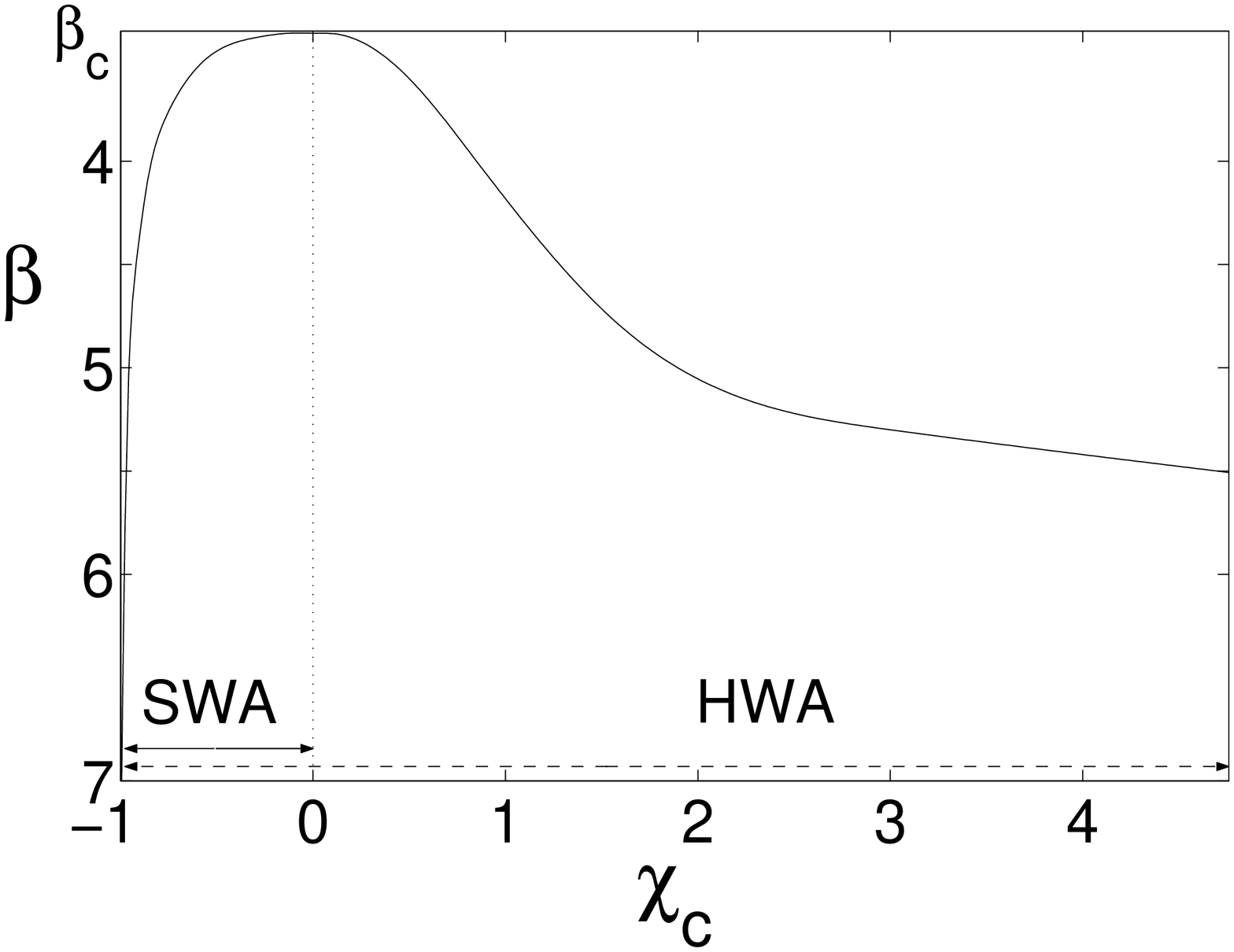}
 \end{center}
 \caption{
The dependence of the critical values of the Hamaker constant $\chi_c$ on inverse temperature
$\beta$ at $\mu=\mu_0$. The left branch  [$\chi_c(\beta)<0$] is the limit $\chi_c^-$ of the
emergence of the precursor layer identical for both SWA and HWA models. The right branch
[$\chi_c(\beta) \ge 0$] denotes the wetting transition to for HWA. The wetting transition for SWA
occurs at $\chi_c=0$ independently of $\beta$.
  }
\label{fig:beta_chic}
\end{figure}

On the other hand, by fixing the value of $\beta$ and varying $\chi$, one finds that a precursor
layer may exist only above a critical value $\chi>\chi_c^-$, as shown in Fig.~\ref{fig:beta_h0}(b).
When the layer thickness is defined by the integral expression (\ref{2inths}), this transition
loses a qualitative character, and the value $\chi_c^-$ can be defined as a point where $\Delta
\mu_d (h_0)=0$ [see Fig.~\ref{fig:slg_H_den}(a)]. This happens at $\chi= \chi_c^-(\beta)
=-(1-\rho^-/\rho^+)$, so that the lower limit is identical for both models (see
Fig.~\ref{fig:beta_chic}). According to the integral formula, negative values are possible, and may
appear when fluid-substrate interactions are so weak that the fluid is nonwetting even at vapor
densities. Moreover, at $\chi$ slightly above $\chi_c^-$, the topology of the curves $\Delta
\mu_d(h)$ for the HWA model changes: the curve becomes discontinuous, and the microscopic and
macroscopic branches of the curve separate, as seen in Fig.~\ref{fig:slg_H_den}(a). The
discontinuity is explained by the absence of the vapor-fluid coexistence above some critical value
of the chemical potential as shown in the inset of Fig.~\ref{fig:slg_H_den}(a). The sequence of
density profiles in the vicinity of $\chi=\chi^-_c$ is shown in Fig.~\ref{fig:slg_H_den}(b).

\subsubsection{Wetting transitions}

A precursor film exists within the interval $\chi_c^-<\chi<\chi_c^+$. The latter limit
corresponds to the wetting transition, as presented in Fig.~\ref{fig:beta_h0}(b). One can see a qualitative difference between the SWA and the HWA models.
For SWA, the precursor thickness grows continuously as $\chi$ increases, and diverges at
$\chi=\chi_c^+=0$, indicating a second-order transition to complete wetting. For HWA, the precursor
layer approaches a finite thickness at finite $\chi=\chi_c^+$, indicating a first order transition
to wetting. The locus of the wetting transition $\chi=\chi_c^+(\beta)$ is determined by numerical
computation (see Fig.~\ref{fig:beta_chic}).

The two limiting cases of SWA and HWA correspond to compliant (fluid or rough solid) or molecularly
smooth solid substrates, respectively. Recently the transitions of both kind (first and second
order) were found in experiments on liquid substrates~\cite{RBM:99,Ra:04}. The quantitative
disagreement between the SWA results discussed here and in the later experiments is due to the
absence of short range interactions in our computations. A more complete picture of wetting
transitions in SWA arises when, in addition to long-range interactions, short-range forces between
the fluid molecules and the substrate are taken into account~\cite{RBM:99,BBSRDPBMI:01}. While
concentrating in the following on short-range interactions in SWA, we refer the reader to
Ref.~\cite{dagama} which describes the impact of short-range interactions in HWA.

\subsection{Short-range interactions in SWA}

\begin{figure}
\begin{center}
  \includegraphics[angle=270,width=3.2in]{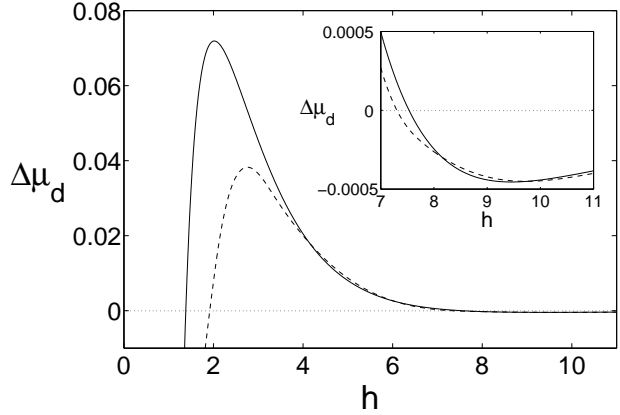}
  \caption{The dependence of the dimensionless disjoining potential $\Delta \mu_d$ on separation $h$.
  The solid line denotes the sharp interface approximation form [Eq.~(\ref{eq:dis_pot})] while the
  direct integration of~(\ref{ffgw_ss}) is demonstrated by the dashed line.
  Parameters: $\beta=9,\,\eta=0.01$ and $\chi=0.3$.
  }
\label{fig:an_num}
\end{center}
\end{figure}

Computation of repulsive short-range interactions in the soft wall approximation takes into account polar
interactions. In the simplest description, the interaction kernel for short-range forces has an
exponential decay~\cite{Lip:84,Shar:93}:
\begin{equation}\label{ff1s_sq_a}
    \psi_s=\pi C_{p} \int_{-\infty}^0 {\rm d} \zeta \int_{q_0}^\infty e^{-\lambda \sqrt{q}} \,{\rm d}
    q\,,
\end{equation}
where $\lambda>0$. Some other forms can be also used here~\cite{ArEv:02}. Unlike the computation of
$\psi_l$, no cutoff is required, and the lower integration limit is $q_0=(z-\zeta)^2$.  This yields
\begin{equation}\label{ff1s_sq_b}
    \psi_s=\frac{2\pi C_{p}}{\lambda^3}\left( 2+\lambda z\right)e^{-\lambda z} \, .
\end{equation}
The modified dimensionless Euler--Lagrange equation reads
\begin{equation}\label{ffgw_ss}
 g(\rho) - \mu+3\eta \beta {\psi}_s(z) +\frac{3}{4} \beta {\psi}_l(z) \left
[\rho^+(\chi+1) -\rho(z)\right ] + \frac{3}{4} \beta \int\limits_{0}^\infty  Q(\zeta-z) [\rho
(\zeta)-\rho (z)] \, {\rm d} \zeta=0
\end{equation}
where for simplicity we set $\lambda=d^{-1}$, ${\psi}_s(z)=(2+z)\exp(-z)$ at $z>0$ and
$\eta=C_{p}b^2/a \sim C_{p}/C_W$. An analytical form of the disjoining potential can be obtained
exploiting the sharp interface approximation for both $\eta$ and $\chi \ll 1$. Adapting the method
discussed in~\cite{pci,pre01} we find
\begin{equation}\label{eq:dis_pot}
    \Delta \mu_d=\frac{\beta}{4} \left [ 12 \eta(2+h)e^{-h}-\rho^+\frac{\chi}{h^3}-\frac{\rho^+ \beta}{8
    g\,'(\rho^+)}\frac{1}{h^6}\right ]\,.
\end{equation}
The derived analytical form of the disjoining potential~(\ref{eq:dis_pot}) is presented in
Fig.~\ref{fig:an_num}. A comparison with direct numerical integration of (\ref{ffgw_ss}) shows a good
agreement already at $h<{\cal O}(10)$ values.

\begin{figure}[htb]
\begin{center}
\includegraphics[angle=270,width=3.4in]{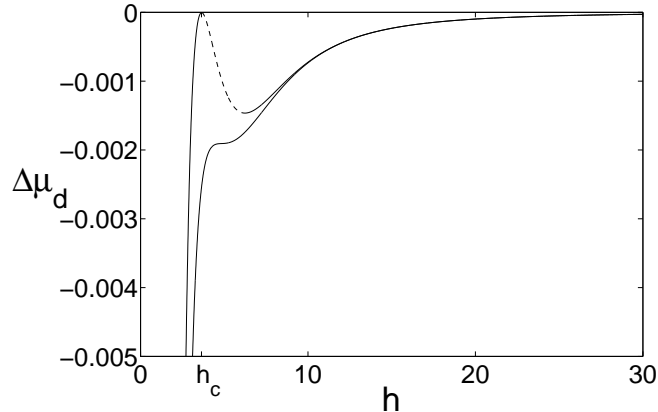}
  \caption{The dependence of the dimensionless precursor layer thickness on the disjoining potential according
  to Eq.~\ref{eq:dis_pot}. The lower line ($\eta=0.00344$) corresponds to the continuous intermediate wetting transition.
  The upper line  ($\eta=\eta_c=0.00407$) corresponds to onset of the frustrated complete wetting. The solid/dashed lines  represent stable/unstable solutions, respectively.
  The Parameters: $\beta=9$ and $\chi=0.4$.
  }
\label{fig:chi_plus}
\end{center}
\end{figure}

A qualitative nature of wetting transitions can be captured via a simple analysis of the disjoining
potential~(\ref{eq:dis_pot}) at the equilibrium chemical potential $\mu=\mu_0$ and thus $\Delta
\mu_d=0$. We distinguish between three characteristic types of behavior which are attributed
to the three corresponding wetting regions:
\begin{description}
    \item[Complete wetting -] there is a single diverging solution ($h \rightarrow
    \infty$) at $\chi>0, \Delta \mu_d=0$, which corresponds to a macroscopic layer.
    \item[Frustrated complete wetting -] three solutions exist for $\chi>0$,
    two of which are finite and the third one is diverging. The two finite solutions have distinct scales,
     molecular (``microscopic'') and mesoscopic.
    \item[Partial wetting -] two solutions exist for $\chi<0$, where one is finite on a molecular scale
    and the other one diverges.
\end{description}
The separating boundary between the partial and frustrated complete wetting depends solely on the
sign change of the Hamaker constant, $\chi$. The transition from frustrated to complete wetting
occurs at a certain critical value of $\eta=\eta_c$ when the maximum of the curve $\Delta \mu_d
(h)$ crosses zero; this happens at a critical thickness $h=h_c$ defined as
\begin{equation}\label{eq:extrm}
    \Delta \mu_d(h_c)=0,~~~~\left.\frac{{\rm d}\Delta \mu_d}{{\rm d}h}\right \vert_{h=h_c}=0,~~~~
    \left.\frac{{\rm d}^2\Delta \mu_d}{{\rm d}h^2}\right \vert_{h=h_c}<0\,.
\end{equation}
The dependence $\Delta \mu_d (h)$ at $\eta=\eta_c$ is shown by the upper curve in Fig.~\ref{fig:chi_plus}. This corresponds to a discontinuous (first order) \textit{intermediate wetting} transition from
microscopic to mesocopic films.
The lower curve with an inflection point in Fig.~\ref{fig:chi_plus}
corresponds to a continuous (second order) \textit{intermediate wetting} transition, which is observed at
a certain shifted value of chemical potential $\Delta \mu_d <0$.

\begin{figure}
\begin{center}
  \includegraphics[angle=270,width=3.4in]{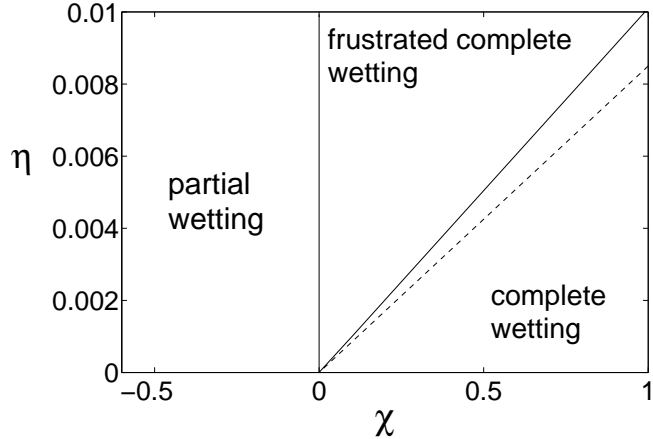}
  \caption{The boundaries of wetting regions (solid line) in the parametric plane $\chi-\eta$ for
  $\beta=9$ and at the equilibrium chemical potential $\Delta \mu_d=0$. The dashed line depicts the
  continuous intermediate  wetting transition at $\Delta \mu_d < 0 $.
  }
\label{fig:chi_A}
\end{center}
\end{figure}

Figure~\ref{fig:chi_A} summarizes these three transitions in the $\chi-\eta$ plane. The critical
point $\chi=\eta=0$ marks the \textit{critical end point}~\cite{Ra:04} and corresponds to the sign
change of the Hamaker constant; all three above wetting regions converge at the critical end point.
Changing the inverse temperature $\beta$ (for constant values of $\eta$) does not change this
picture in a qualitative way.

\section{Discussion}

In the current study we have presented an analysis of wetting phenomena and interactions between
liquid-vapor interfaces at microscopic, mesoscopic and macroscopic distances based on a relatively
simple local density functional theory (DFT). The study is centered on computation of disjoining
and conjoining potentials (which are important for dynamic computations). Focusing on the role of
long-range van der Walls interactions, we considered two models -- hard wall and soft wall
approximations -- differing by the role of steric effects. The first difference corresponds to a
precursor layer thickness and the quantitative form of the disjoining potential. Under similar
external conditions we found that droplets on smooth surfaces will roll-up before the ones on
complaint surfaces~\cite{OSST:96,MSAPN:04} due to the steric forces. The distinct wall models also
lead to a qualitatively different character of wetting transitions. A first order transition from
partial to complete wetting occurs in the hard wall model only, turning to a second order
transition in the soft wall model.

To capture a more realistic description of wetting transitions, we included a weak dependence on
short-range polar interactions. Analytical derivation of the disjoining potential was carried out
exploiting the sharp interface approximation, which showed a good agreement with a direct numerical
solution of DFT Euler-Lagrange integral equations. We find that the qualitative nature of the
sequence of  wetting transitions stems from a competition between short- and long-scale
interactions, best seen in the ($\chi$, $\eta$) parametric plane, while all other parameters, like
temperature $\beta^{-1}$ or short-range decay range $\lambda$ have only a quantitative effect. The
role of repulsive short-range interactions is analogous to steric effects, which are emphasized in
the hard wall approximation. Thus, the presented simple model reproduces all major types of wetting
transitions (between partial to complete/frustrating wetting and thin/thick precursor in
coexistence with bulk fluid) of~\cite{SBRBM:98,RBM:99,BBSRDPBMI:01,Ra:04} as summarized in
Fig.~\ref{fig:chi_A}.

\subsection*{Acknowledgement}
This research has been supported by Israel Science Foundation.


\begin{thebibliography}{40}

\bibitem{pp} L.M.\ Pismen and Y.\ Pomeau, Phys. Rev. E, 62 (2000) 2480.
\bibitem{pci} L.M.\ Pismen, Colloids Surfaces A: Physicochem. Eng. Aspects, 206 (2002) 11.
\bibitem{pom02} Y.\ Pomeau,  C.R.\ Mecanique, 330 (2002) 207.
\bibitem{DGen85} P.G.\ de Gennes, Rev. Mod. Phys., 57 (1985) 827.
\bibitem{phf04} L.M.\ Pismen and Y.\ Pomeau, Phys. Fluids, 6 (2004) 2604.
\bibitem{ebner} C.\ Ebner and W.F.\ Saam, Phys. Rev. Lett., 38 (1977) 1486.
\bibitem{NaFi:82} H.\ Nakanishi and M.E.\ Fisher, Phys. Rev. Lett., 49 (1982) 1565.
\bibitem{LiKr:84} R.\ Lipowsky and D.M.\ Kroll, Phys. Rev. Lett., 52 (1984) 2303.
\bibitem{DiSc:85} S.\ Dietrich and M.\ Schick, Phys. Rev. B, 31 (1985) 4718.
\bibitem{RuTa:92} J.E.\ Rutledge and P.\ Taborek, Phys. Rev. Lett., 69 (1992) 937.
\bibitem{BKW:92} D.\ Bonn, H.\ Kellay, and G.H.\ Wegdam, Phys. Rev. Lett., 69 (1992) 1975.
\bibitem{KBM:93} H.\ Kellay, D.\ Bonn, and J.\ Meunier, Phys. Rev. Lett., 71 (1993) 2607.
\bibitem{RMBIB:96} K.\ Ragil, J.\ Meunier, D.\ Broseta, J.O.\ Indekeu, and D.\ Bonn, Phys. Rev. Lett., 77 (1996) 1532.
\bibitem{SBRBM:98} N.\ Shahidzadeh, D.\ Bonn, K.\ Ragil, D.\ Broseta, and J.\ Meunier, Phys. Rev. Lett., 80 (1998) 3992.
\bibitem{RBM:99} D.\ Ross, D.\ Bonn, and J.\ Meunier, Nature, 400 (1999) 737.
\bibitem{BBSRDPBMI:01} D.\ Bonn, E.\ Bertrand, N.\ Shahidzadeh, K.\ Ragil, H.T.\ Dobbs, A.I.\ Posazhennikova, D.\ Broseta, J.\ Meunier, and J.O.\ Indekeu, J. Phys.: Condens. Mat., 13 (2001) 4903.
\bibitem{der}  B.V.\ Derjaguin, N.V.\ Churaev and V.M.\ Muller, Surface Forces, Consultants Bureau, New York, 1987.
\bibitem{cusp} L.M.\ Pismen, Phys. Rev. E, 70 (2004) 51604.
\bibitem{vdw} J.D.\ van der Waals (English translation: J.S.\ Rowlinson), J.\ Stat.\ Phys., 20 (1979) 197.
\bibitem{Eva:79} R.\ Evans, Adv. Phys., 28 (1979) 143.
\bibitem{ditr} S.\ Dietrich and M. Napi\`orkowski, Phys. Rev. A, 43 (1991) 1861.
\bibitem{dagama} A.\ Gonz\'{a}lez and M.M.\ Telo da Gama, Phys. Rev. E, 62 (2000) 6571.
\bibitem{OSST:96} T.\ Onda, S.\ Shibuichi, N.\ Satoh, and K.\ Tsujii, Langmuir 12 (1996) 2125.
\bibitem{MSAPN:04} G.\ McHale, N.J.\ Shirtcliffe, S.\ Aqil, C.C.\ Perry, and M.I.\ Newton, Phys. Rev. Lett., 93
(2004) 036102.
\bibitem{ll} L.D.\ Landau and E.M.\ Lifshitz, v. V, Statistical Physics, Part I, Pergamon Press, 1980.
\bibitem{pre01} L.M.\ Pismen, Phys.\ Rev. E, 64 (2001) 021603.
\bibitem{henderson} J.R.\ Henderson and Z.A.\ Sabeur, J. Chem. Phys., 97 (1992) 6750.
\bibitem{Ra:04} S.\ Rafa\"{\i}, D.\ Bonn, E.\ Bertrand, J.\ Meunier, V.C.\ Weiss, and J.O.\ Indekeu, Phys. Rev. Lett., 92 (2004) 245701.
\bibitem{Lip:84} R.\ Lipowsky, Phys. Rev. Lett., 52 (1984) 1429.
\bibitem{Shar:93} A.\ Sharma, Langmuir, 9 (1993) 861.
\bibitem{ArEv:02} A.J. Archer and R. Evans, J. Phys.: Condens. Matter, 14 (2002) 1131.
\end{thebibliography}
\end{document}